# Advanced Statistical Methods for Large Scale Structure Studies

Stefano Borgani

INFN–Sezione di Perugia, c/o Dipartimento di Fisica dell'Universitá, via A.Pascoli, I-06100 Perugia, Italy
SISSA, International School for Advanced Studies, via Beirut 2–4, I–34014 Trieste, Italy

## 1. Introduction

The aim of these *Lectures* is to provide an overview of statistical tools, which are currently used for the study of the large–scale distribution of cosmic structures and which go beyond the simple (although useful!) two–point correlation function. The reason why we need such "higher–order" informations lies essentially in the fact that two–point correlations exhaust the statistical content of a system only in the case it has Gaussian nature. On the other hand, even allowing for random–phase initial conditions, there are good reasons to expect that the present–day distribution of galaxies and galaxy clusters have non–Gaussian features, which are rooted in the dynamical history of their formation and evolution.

It is clear that a statistical description of the large–scale texture of the Universe, which is as complete as possible, is not only required to provide a cosmographical description, i.e. to merely understand whether galaxies are preferentially located in clusters or in filaments, or by how much they leave devoid the underdense parts of their distribution. Instead, this information is a necessary ingredient to solve the dynamical problem of cosmic structure formation: once an assumption is made for the underlying dynamics governing the evolution (e.g., gravitational instability), and having (almost) fixed the amplitude of large–scale fluctuations thanks to COBE, one's hope is that the statistical knowledge of the galaxy distribution at the present time should be univocally related to the nature of primordial fluctuations and, hopefully, to the dark matter content of the Universe.

Having this in mind, it is clear that the characteristics we would desire for statistical descriptors are the following (see also ref.[64]):

**(a)** to be **robust**, that is to be able to provide statistically significant results even when dealing, as usual, with rather limited data sets;

**(b)** to be **discriminative**, so as to pick up significant differences when applied to different dark–matter models;

**(c)** to be **interpretable**, so that the statistical information it provides can be easily connected to dynamical and physical quantities;

**(d)** to be **assumption–free**, so that the results it provides are not sensitive to the way of identifying galaxies and galaxy clusters in collisionless simulations.

After the first attempts to statistically describe the galaxy clustering, which date back to about 25 years ago, many such methods have been proposed and applied. Providing a description for all of them would go by far beyond the scope of these *Lectures*. Therefore, in the following I will mainly deal with those of such methods which have been more commonly used. Furthermore, instead of providing technical details about their implementation in practical applications, I will discuss which kind of information they provide and what we have learned up to now from them. For readers which were interested in specific aspect, as well as on other statistical methods, I refer to refs.[8, 69] for recent comprehensive reviews, as well as to the relevant literature quoted therein.

In more detail, I will concentrate on correlation analysis methods and related issues, like count–in–cell statistics and probability density function. This choice is not just motivated by historical reasons (correlation functions have been the first quantities which have been measured in extended galaxy samples), but mainly by the fact that this type of analysis is still today the most commonly applied to analyze both real data sets and numerical simulations. In this context, Section 2 introduces the statistical formalism, while Section 3 is devoted to a brief description of the main sources of uncertainties in statistical analyses. Section 4 describes the results of correlation analysis of observational samples and their interpretation. Section 5 deals with geometrical descriptions of the large–scale clustering, like those provided by the void probability function and topological characteristics. Final comments are deserved to Section 5.

## 2. The correlation statistics

The classical correlation analysis of the galaxy distribution was based on the determination of the 2–point correlation function, $\xi(r)$. Its definition is related to the joint probability

$$\delta^{(2)}P \;=\; n^2\,\delta V_1\,\delta V_2\,[1+\xi(r_{12})] \tag{1}$$

of finding an object in the volume element $\delta V_1$ and another one in $\delta V_2$, at separation $r_{12}$. In eq.(1) the factorization of the $n^2$ term ($n$ being the galaxy mean number density) makes $\xi(r)$ a dimensionless quantity and the total probability turns out to be normalized to the square of the total number of object in the distribution. According to its definition, the value of the correlation function is a measure of the non–random behaviour of the distribution and, for an isotropic clustering, depends only on the modulus of the separation vector $\mathbf{r}_{12}$. In particular, object positions are said to be correlated if $\xi(r) > 0$ and anticorrelated if $-1 \le \xi(r) < 0$, while a Poissonian distribution is characterized by $\xi(r) = 0$ at any separation.

The concept of correlation functions can be extended to higher orders, by considering the joint probabilities between more than two points. In the following I will introduce the concept of correlations of generic order for a given density field.

### *2.1. Correlation functions*

Let us consider a generic density field, $\rho(\mathbf{x})$, and the relative fluctuations, $\delta(\mathbf{x}) = (\rho(\mathbf{x}) - \bar{\rho})/\bar{\rho}$ around the average density $\bar{\rho}$. By definition, it is $\langle\delta(\mathbf{x})\rangle = 0$, while the requirement of a positively defined $\rho(\mathbf{x})$ leads to $\delta(\mathbf{x}) > -1$. In the following, $\delta(\mathbf{x})$ is assumed to be described by a random function, so that the Universe can be considered as a particular realization taken from an ensemble (functional space) $\mathcal{F}$ containing all the $\delta(\mathbf{x})$ fields satisfying the above two requirements.

In order to describe the statistics of the $\delta(\mathbf{x})$ field, let $\mathcal{P}[\delta(\mathbf{x})]$ be the probability that the density fluctuations are described by a given $\delta(\mathbf{x}) \in \mathcal{F}$. With the assumption of statistical homogeneity, the probability functional $\mathcal{P}[\delta(\mathbf{x})]$ turns out to be independent of the position $\mathbf{x}$, while, due to the requirement of isotropic clustering, the joint distribution of $\delta(\mathbf{x}_1)$ and $\delta(\mathbf{x}_2)$ depends only on the the separation $r_{12} = |\mathbf{x}_1 - \mathbf{x}_2|$. By definition, the probability distribution in the functional space must be normalized so that the total probability is unity: $\int_{\mathcal{F}} \mathcal{D}[\delta(\mathbf{x})]\, \mathcal{P}[\delta(\mathbf{x})] = 1$. Here $\mathcal{D}[\delta(\mathbf{x})]$ represents a suitable measure introduced in $\mathcal{F}$ in order to define the functional integral.

Let us consider the *partition functional*

$$\mathcal{M}[\tau(\mathbf{x})] \equiv \int \mathcal{D}[\delta(\mathbf{x})]\, \mathcal{P}[\delta(\mathbf{x})]\, e^{\int d\mathbf{x}\, \delta(\mathbf{x})\tau(\mathbf{x})} = \langle e^{\int d\mathbf{x}\, \delta(\mathbf{x})\tau(\mathbf{x})} \rangle , \quad (2)$$

where $\tau(\mathbf{x})$ is a generic function, that plays the role of an external source perturbing the underlying statistics. A complete characterization of the statistics of the density distribution can be given in terms of the $n$–point correlation functions

$$\mu_n(\mathbf{x}_1, \ldots, \mathbf{x}_n) = \langle \delta(\mathbf{x}_1) \ldots \delta(\mathbf{x}_n) \rangle = \left. \frac{\delta^n \mathcal{M}[\tau]}{\delta\tau(\mathbf{x}_1) \ldots \delta\tau(\mathbf{x}_n)} \right|_{\tau=0} , \quad (3)$$

with $\mu_1(\mathbf{x}) = \langle \delta(\mathbf{x}) \rangle = 0$. The notation $\langle \cdot \rangle$ indicates the average over the $\mathcal{F}$ space while $\delta/\delta\tau(\mathbf{x})$ stands for the functional derivative with respect to $\tau(\mathbf{x}) \in \mathcal{F}$ (see Appendix for the meaning of the functional derivative). Eq.(3) represents the statistical–mechanical equivalent of the path–integral definition of the Green's functions in quantum field theory (see, e.g., ref.[62]). Under the assumption of *ergodicity* of our system, the averages taken over the (physical) configuration space are completely equivalent to the expectations taken over an ensemble of universes, i.e. over the functional space $\mathcal{F}$. From now on I will indifferently use the symbol $\langle \cdot \rangle$ to indicate both kinds of average.

A further characterization of $\delta(\mathbf{x})$ is also given in terms of the *connected* or *irreducible* correlation functions, $\kappa_n(\mathbf{x}_1, \ldots, \mathbf{x}_n)$. Such quantities are introduced through their generating functional

$$\mathcal{K}[\tau(\mathbf{x})] \equiv \ln \mathcal{M}[\tau(\mathbf{x})] , \quad (4)$$

so that

$$\kappa_n(\mathbf{x}_1, \ldots, \mathbf{x}_n) = \left. \frac{\delta^n \mathcal{K}[\tau]}{\delta\tau(\mathbf{x}_1) \ldots \delta\tau(\mathbf{x}_n)} \right|_{\tau=0} . \quad (5)$$

Therefore, a unique characterization of the statistics, i.e. the knowledge of the partition functions, requires that correlation functions of any order are known.

For $n = 2$, it is easy to show that the definition (3) of correlation function is completely equivalent to that provided by eq.(1). In fact, the 2–point joint probability of having the density values $\rho(\mathbf{x}_1)$ in the position $\mathbf{x}_1$ and $\rho(\mathbf{x}_2)$ in $\mathbf{x}_2$ is $\langle \rho(\mathbf{x}_1)\, \rho(\mathbf{x}_2) \rangle = \bar{\rho}^2 [1 + \mu_{2,12}]$, which coincides with eq.(1), once we take $\xi(r_{12}) = \mu_2(r_{12})$.

In order to study the structure of the 3–point correlation function, let us suppose that the point $\mathbf{x}_3$ is sufficiently far away from $\mathbf{x}_1$ and $\mathbf{x}_2$, so that the event probability in $\mathbf{x}_3$ does not depend on that in the other two points. If this is the case, the 3–point joint probability is

$$\langle \rho_1\, \rho_2\, \rho_3 \rangle = \langle \rho_1\, \rho_2 \rangle \times \bar{\rho} , \quad (6)$$

where the meaning of the indices is obvious. Hence, requiring symmetry for the exchange of $\mathbf{x}_3$ with $\mathbf{x}_1$ and with $\mathbf{x}_2$, the 3–point probability can be cast in the form

$$\langle \rho_1 \, \rho_2 \, \rho_3 \rangle \; = \; \bar{\rho}^3 \left[1 + \xi_{12} + \xi_{23} + \xi_{13} + \zeta_{123}\right] \quad . \tag{7}$$

Here, $\zeta \equiv \kappa_3$ is the term that correlates the three points all together and must vanish when one of these points is removed:

$$\zeta(\mathbf{x}_i, \mathbf{x}_j, \mathbf{x}_l \to \infty) \; = \; 0 \qquad i \neq j \neq l \quad ; \quad i, j, l = 1, 2, 3 \, . \tag{8}$$

On the basis of similar considerations, the 4–point joint probability is written as

$$\langle \rho_1 \, \rho_2 \, \rho_3 \, \rho_4 \rangle \; = \; \bar{\rho}^4 \left\{1 + \left[\xi_{12} + ...6 \text{ terms}\right] + \left[\zeta_{123} + ...4 \text{ terms}\right] + \mu_4\right\} . \tag{9}$$

Here the 4–point correlation function

$$\mu_{4,1234} \; = \; \xi_{12}\xi_{34} + \xi_{23}\xi_{14} + \xi_{13}\xi_{24} + \eta_{1234} \tag{10}$$

represents the term connecting the four points and gives a vanishing contribution when at least one point is moved to infinite separation from the others. The $\mu_4$ term contains three terms connecting two pairs separately, while $\eta \equiv \kappa_4$ is the usual notation to indicate the connected 4–point function, which accounts for the amount of correlation due to the simultaneous presence of the four points.

In general, correlations of order $n$ are related to the $n$–point joint probability,

$$\langle \rho(\mathbf{x}_1) \ldots \rho(\mathbf{x}_n) \rangle \; = \; \bar{\rho}^n \left[1 + \left(\text{terms of order} < n\right) + \mu_n(\mathbf{x}_1, ..., \mathbf{x}_n)\right] , \tag{11}$$

in such a way that they give null contribution when any subset of $\{\mathbf{x}_1, \ldots, \mathbf{x}_n\}$ is removed to infinity. In turn, an important theorem of combinatorial analysis shows that, removing from the $\mu_n$ function all the disconnected contributions, the remaining connected part is just the $\kappa_n$ function defined by eq.(5). The general proof of this theorem is rather tricky and will not be reported here. It is however not difficult to see that, expressing the derivatives of the $\mathcal{K}[\tau]$ partition function in terms of that of $\mathcal{M}[\tau]$, we get at the first correlation orders

$$\mu_2 = \kappa_2 \, , \qquad \mu_3 = \kappa_3 \, , \qquad \mu_4 = 3\kappa_2^2 + \kappa_4 \, , \qquad \mu_5 = 10\kappa_2\kappa_3 + \kappa_5 \, ,$$
$$\mu_6 = 15\kappa_2^3 + 10\kappa_3^2 + 15\kappa_2\kappa_4 + \kappa_6 \, . \tag{12}$$

From eq.(11), it follows that the $n$–point functions measure by how much the distribution differs from a completely random (Poissonian) process. In fact, for a Poissonian distribution the probability of some events in any subset of $\{\mathbf{x}_1, ..., \mathbf{x}_n\}$ does not affect the probability in the other points. Accordingly, $\langle \rho(\mathbf{x}_1) \ldots \rho(\mathbf{x}_n) \rangle = \bar{\rho}^n$ and correlations of any order vanish.

### *2.2. Correlations of a Gaussian field*

A particularly interesting and simple case is that in which the density fluctuations are approximated by a random Gaussian process. The important role of Gaussian perturbations in cosmological context lies in the fact that, according to the classical inflationary scenario, they are expected to be originated from quantum fluctuations of a scalar field at the outset of the inflationary expansion (see, e.g., ref.[57] and references therein). Even without resorting to inflation, the Central Limit Theorem guarantees that the Gaussian statistics is the consequence of a large variety of random processes, which makes it a sort of natural choice.

The Gaussian probability distribution in the functional space $\mathcal{F}$ takes the form

$$\mathcal{P}[\delta(\mathbf{x})] \;=\; (\det C)^{-1/2} \exp\left\{-\frac{1}{2}\int d\mathbf{x} \int d\mathbf{x}'\,\delta(\mathbf{x})C^{-1}(\mathbf{x},\mathbf{x}')\delta(\mathbf{x}')\right\} . \quad (13)$$

Here $C(\mathbf{x},\mathbf{x}')$ is called the *correlation operator*, which must be invertible and symmetric with respect to the variables $\mathbf{x},\mathbf{x}'$. From eq.(13), it follows that this operator determines the variance of the distribution and, more generally, the correlation properties of the fluctuation field. The above expression of the probability distribution is such as to satisfy the normalization requirement. The corresponding partition functional $\mathcal{M}[\tau]$ is

$$\begin{aligned}\mathcal{M}[\tau] = (\det C)^{-1/2} \int \mathcal{D}[\delta(\mathbf{x})] \, \exp\Big[-\frac{1}{2}\int d\mathbf{x} \int d\mathbf{x}'\,\delta(\mathbf{x})C^{-1}\delta(\mathbf{x}') + \\ + \int d\mathbf{x}\,\delta(\mathbf{x})\,\tau(\mathbf{x})\Big] \\ = \exp\left[\frac{1}{2}\int d\mathbf{x} \int d\mathbf{x}'\,\tau(\mathbf{x})C\tau(\mathbf{x}')\right] . \quad (14)\end{aligned}$$

According to the definition (4) of $\mathcal{K}[\tau]$, the generator of the connected correlation functions reads

$$\mathcal{K}[\tau] \;=\; \frac{1}{2}\int d\mathbf{x} \int d\mathbf{x}'\tau(\mathbf{x})C(\mathbf{x},\mathbf{x}')\tau(\mathbf{x}')\,, \quad (15)$$

so that the corresponding connected correlation functions are

$$\begin{aligned}&\kappa_2(\mathbf{x}_1,\mathbf{x}_2) = C(\mathbf{x}_1,\mathbf{x}_2)\\ &\kappa_n(x_1,...,x_n) \;=\; 0 \qquad\qquad \text{if} \quad n > 2\,. \quad (16)\end{aligned}$$

Therefore, the fundamental property of a Gaussian density field is that its statistics is completely determined by 2–point correlations.

Although Gaussian density fluctuations are the natural outcome of simple inflationary schemes, nevertheless the observed distribution of cosmic structures displays a clear non–Gaussian behaviour, as the detection of non–vanishing higher–order correlations shows (see below). However, even starting with an initial Gaussian density field, there are at least two valid motivations to understand the development of subsequent non–Gaussian statistics for the galaxy distribution. Firstly, note that the Gaussian statistics assign a non vanishing probability even to the unphysical values $\delta(\mathbf{x}) < -1$. However, as long as the variance of $\delta$ is much less than unity, the Gaussian distribution is a good approximation, since a negligible probability is assigned to $\delta < -1$. On the other hand, as soon as $\delta$ grows by gravitational instability, it is allowed to keep arbitrarily large and positive values while the $\delta < -1$ region remains forbidden, thus forcing $\mathcal{P}[\delta]$ to become more and more skewed. Secondly, non–Gaussian statistics is also expected in the framework of "biassed" models of galaxy formation [39, 3], in which the observed cosmic structures are identified with those peaks of the underlying Gaussian matter field that exceed a critical density value. In this case, analytical argument [63, 36] shows that the non–Gaussian behaviour arises as a threshold effect superimposed on a Gaussian background.

*2.3. The count–in–cell statistics*

In order to pass from the study of a continuous density field to that of a single variable, let us suppose to sample the density field $\rho(\mathbf{x})$ with volume elements of size $R$, whose shape is described by the window function $W_R(\mathbf{x})$. The resulting observable quantity is the variable

$$\rho_R \;=\; \int d^3x\, \rho(\mathbf{x})\, W_R(\mathbf{x})\,. \tag{17}$$

The function $W_R(\mathbf{x})$ is normalized so that $\int d^3xW(\mathbf{x}) = 1$ and $\rho_R$ is the local average of the density within the sampling volume. Commonly adopted choices for $W_R$ are the top–hat window,

$$W_R(\mathbf{x}) \;=\; \left(\frac{4\pi}{3}R^3\right)^{-1} \theta\left(1-\frac{|\mathbf{x}|}{R}\right) \tag{18}$$

and the Gaussian window

$$W_R(\mathbf{x}) \;=\; (2\pi R^2)^{-3/2} e^{-|\mathbf{x}|^2/2R^2}\,. \tag{19}$$

In analogy with the correlation function description of the continuous field $\rho(\mathbf{x})$, the statistics of $\rho_R$ is completely specified by its probability density function (PDF) $p(\rho_R)$. The corresponding moment of order $n$ is

$$m_n(R) \;=\; \int d\rho_R\, p(\rho_R) \left(\frac{\rho_R}{\bar{\rho}}\right)^n\,. \tag{20}$$

The moment generating function (MGF) is defined as

$$M(t) \;=\; \langle\, \exp\left(t\rho_R/\bar{\rho}\right) \rangle \;=\; \int d\rho_R\, p(\rho_R)\, e^{t\rho_R/\bar{\rho}} \tag{21}$$

and is the analogous of the $\mathcal{M}(\tau)$ functional of the continuous case. In turn, the MGF can be expanded in McLaurin series,

$$M(t) \;=\; \sum_{n=0}^{\infty} \frac{m_n(R)}{n!}\, t^n \qquad ; \qquad m_n(R) \;=\; \left.\frac{d^n M(t)}{dt^n}\right|_{t=0}\,. \tag{22}$$

where the moments $m_n(R)$ are related to the correlation functions $\mu_n$ though the integral relation

$$m_n(R) = \int d^3x_1 \ldots \int d^3x_n\, W_R(\mathbf{x}_1) \ldots W_R(\mathbf{x}_n) \times \\ \left[1 + \left(\text{terms of order} < n\right) + \mu_n(\mathbf{x}_1, ..., \mathbf{x}_n)\right]\,. \tag{23}$$

In a similar fashion, the cumulants or irreducible moments $k_n(R)$ are defined through the generating function

$$K(t) \;\equiv\; \ln M(t) \;=\; \sum_{n=0}^{\infty} \frac{k_n}{n!}\, t^n \qquad ; \qquad k_n \;\equiv\; \left.\frac{d^n K(t)}{dt^n}\right|_{t=0}\,, \tag{24}$$

which is analogous to the $\mathcal{K}[\tau]$ generator of connected correlations. In fact, the cumulant turns out to be related to the connected functions according to

$$k_n(R) \;=\; \int d^3x_1 \ldots \int d^3x_n\; W_R(\mathbf{x}_1) \ldots W_R(\mathbf{x}_n)\, \kappa_n(\mathbf{x}_1, ..., \mathbf{x}_n)\,, \tag{25}$$

which is the average value of the irreducible $n$-point correlation function. Accordingly, $k_2 = \bar{\xi}$ is the variance of the distribution, while $k_3$ is the skewness (see ref.[15] for the relevance of skewness in cosmological context) and $k_4$ is the kurtosis. Suitable relations between $k_n$ and $m_n$ can be found by successively differentiating eq.(24), which resembles the analogous relations between connected and disconnected correlation functions (see eq.[12]).

From eq. (21), the PDF is expressed as the inverse transform of the MGF as

$$p(\rho_R) \;=\; \frac{1}{2\pi i\bar{\rho}} \int_{-i\infty}^{+i\infty} dt\, M(t)\, e^{-it\rho_R/\bar{\rho}} \,. \tag{26}$$

It is worth noting that, for some models of the PDF, although the moments are well defined according to eq.(20), the series in eq.(22) or, equivalently, the integral in eq.(21), do not converge. For instance, this is the case of the lognormal PDF (see Section 4.2, below; see also refs.[16, 17] for a more detailed discussion on this point). In fact, for this model the divergence of the integral in eq.(21) is due to the long high–density tail of the corresponding PDF shape. What happens in cases like this is that the moments of integer positive order do not exhaust the whole one–point statistical information which is contained in the PDF. Obviously, this does not imply that, even for such PDFs, the moments $m_n$ are of scarce relevance. Instead, they are anyway useful instruments to compare data and simulations in order to assess the reliability of cosmological models.

Instead of dealing with continuous distributions, in the analysis of galaxy catalogues as well as of N-body simulations one considers discrete point distributions. Therefore, a suitable prescription is required in order to relate the statistics of the underlying density field to that of its discrete realization. A usual assumption is that the point distribution one deals with represents a Poissonian realization of an underlying continuous field. Let $\bar{N}$ be the average number of objects within the sampling volumes. In a random realization of a given value of $\rho_R$, the actual number of points must obviously be an integer. Its expectation (non–integer) value over all the random realizations of $\rho_R$ is $(\bar{N}/\bar{\rho})\rho_R$, fluctuations around this value being described by a Poissonian statistics. The PDF for a Poisson process $\varphi$ with mean $\bar{\varphi}$ is $p_P(\varphi) = \sum_{N=0}^{\infty} \frac{\bar{\varphi}^N}{N!} e^{-\bar{\varphi}} \delta_D(\varphi - N)$. Therefore, the PDF for a process $x = \varphi\bar{\rho}/\bar{N}$ is $p_P(x) = p_P(\varphi)\bar{N}/\bar{\rho}$. Accordingly, the MGF reads

$$M_P(t) \;=\; \int dx\, p_P(x)\, e^{tx/\bar{\rho}} \;=\; \exp\left[\frac{\bar{N}}{\bar{\rho}}\,\rho_R\,\left(e^{t/\bar{N}} - 1\right)\right] \,. \tag{27}$$

This procedure concerning a particular $\rho_R$ is to be averaged over all the possible realizations of the $\rho_R$ process. In this way we obtain the MGF for the discrete counts, which reads

$$\begin{aligned} M_{disc}(t) &= \int d\rho_R\, p(\rho_R)\, \exp\left[\frac{\bar{N}}{\bar{\rho}}\,\rho_R\,\left(e^{t/\bar{N}} - 1\right)\right] \\ &= M\left[\bar{N}\left(e^{t/\bar{N}} - 1\right)\right] \,. \end{aligned} \tag{28}$$

The discrete nature of the point distribution is therefore accounted for by the change of variable $t \to \bar{N}(e^{t/\bar{N}} - 1)$ in the functional dependence of $M(t)$, which leaves the variable unchanged in the limit $\bar{N} \to \infty$.

As for the PDF, in the discrete case eq.(26) gives

$$p(\rho_R) \;=\; \frac{1}{2\pi\bar\rho}\int_{-\infty}^{+\infty} dt\, M[\bar N(e^{it/\bar N}-1)]\, e^{t\rho_R/\bar\rho}\,. \tag{29}$$

Since the variable $e^{it/\bar N}$ takes values only on the unit circle of the complex plane, the MGF turns out to be a periodic function. Therefore, its Fourier transform can be written as a sum of Dirac $\delta$-functions:

$$p(\rho_R) \;=\; \frac{\bar N}{\bar\rho}\sum_{N=-\infty}^{+\infty}\delta_D\left(\frac{\rho_R}{\bar\rho}-\frac{N}{\bar N}\right)P_N(R)\,. \tag{30}$$

Accordingly, the PDF vanishes except for a discrete set of values of $\rho_R/\bar\rho$, as it must for a point distribution. In the above expression, the coefficients $P_N(R)$ are

$$P_N(R) \;=\; \frac{1}{2\pi i}\oint dy\, y^{-(N-1)} M[\bar N(y-1)]\,. \tag{31}$$

For analytical $M(t)$ all the $P_N$'s for $N<0$ vanish, so that they acquire the meaning of probabilities of finding $N$ points inside a volume of size $R$. For $N\to\infty$ and $\bar N\to\infty$, with fixed $N/\bar N$, eq.(30) gives back the continuous limit $P_N(R)=(\bar\rho/\bar N)p(\rho_R)$, with the effective density variable given by $\rho_R/\bar\rho = N/\bar N$.

The statistics of the point distribution can be described in terms of the central moments $\mu_n = \langle (N-\bar N)^n\rangle/\bar N^n$, where the moments of counts

$$\langle N^n\rangle \;=\; \sum_{N=1}^{\infty} P_N\, N^n \;=\; \left.\frac{d^n M_{disc}(t)}{dt^n}\right|_{t=0} \tag{32}$$

are the coefficients of the McLaurin expansion of the discrete MGF. According to the above relations and following the definition (25) of cumulants, it is possible to express $k_n$, which characterize the underlying continuous field, in terms of the the measured moments of discrete counts. At the lowest orders, it is

$$\begin{aligned}
\mu_2 &= \frac{1}{\bar N} + k_2\,,\\
\mu_3 &= -\frac{2}{\bar N^2} + 3\frac{\mu_2}{\bar N} + k_3\,,\\
\mu_4 &= \frac{6}{\bar N^3} - 11\frac{\mu_2}{\bar N^2} + \frac{\mu_3}{\bar N} + 3\mu_2^2 + k_4
\end{aligned} \tag{33}$$

(see, e.g., ref.[7]), while more cumbersome relations hold at higher orders. As expected, all the shot–noise corrections vanish for large $\bar N$ values, while they dominate the signal when the sampling rate is very low ($\bar N \ll 1$). In this case, recovering the continuous statistics become a rather noisy procedure. It can be also shown that, although the relations (33) have been obtained on the ground of the relation (28) between discrete and continuous MGFs, their validity is not conditioned by the existence of $M(t)$ (see, e.g., ref.[58] for a derivation of shot–noise corrections under general conditions).

A more serious reason of concern in the application of eqs.(33) comes from the fact that they are based on the assumption that the point distribution represents a random sampling of an underlying continuous field. However, if observable objects trace the high–density peaks, as expected for galaxies and clusters of galaxies, they are far from being a Poissonian sampling of the dark matter distribution. Therefore, such corrections are not expected to recover the statistics of the background density field. Furthermore, for generic non–linear relations between density field and object distribution it is not guaranteed a priori the possibility of self–consistently defining a continuous field, for which the observed galaxy distribution represents a Poissonian sampling. For these reasons, it is not necessarily recommendable to apply shot–noise corrections when comparing moments obtained from real data and numerical simulations, once care is taken to reproduce in the artificial data set the same galaxy number density as in the real one.

### *2.4. The hierarchical model*

A rather popular model for connected correlations is represented by the hierarchical ansatz

$$\kappa_n(\mathbf{x}_1,\ldots,\mathbf{x}_n) \;=\; \sum_{n-trees\ a}^{t_n} Q_{n,a} \sum_{labelings} \prod_{edges}^{(n-1)} \kappa_{2,ij}\,, \tag{34}$$

which expresses the $n$–point connected function in terms of products of $(n-1)$ 2–point functions [25]. In eq.(34), distinct "trees" designated by $a$ have in general different coefficients $Q_a$, while the complete sequence of these coefficients uniquely specifies the hierarchical model. Configurations that differ only in interchange of labels 1,...,$n$ all have the same amplitude coefficients, and $ij$ is a single index which identifies links. The number of trees $t_n$ with $n$ vertices is fixed by a theorem of combinatorial analysis, while the total number of labeled trees is $T_n = n^{n-2}$. Thus, eq.(34) has $t_n$ amplitude coefficients $(a = 1,...,t_n)$ and $T_n$ total terms. For instance, for $n = 3$ it is $t_3 = 1$ and $T_3 = 3$, so that

$$\zeta_{123} \;=\; Q\,[\xi_{12}\xi_{13} + \xi_{12}\xi_{23} + \xi_{13}\xi_{23}]\;. \tag{35}$$

For $n = 4$ it is $t_4 = 2$ and $T_4 = 16$. The resulting structure of the 4–point function is

$$\begin{aligned}\eta_{1234} = {} & Q_{4,1}\,[\xi_{12}\xi_{23}\xi_{34} + \ldots (4\text{ terms})] + \\ & Q_{4,2}\,[\xi_{12}\xi_{13}\xi_{14} + \ldots (12\text{ terms})]\;.\end{aligned} \tag{36}$$

By inserting the expression of eq.(34) for the connected correlations into the definition (25) of cumulants, they can be written as

$$k_n \;=\; S_n \bar{\xi}^{n-1}\,, \tag{37}$$

The reduced cumulants $S_n$ are given by suitable combinations of the $Q_{n,a}$ coefficients and their value also depends on the window function profile. Accordingly, the cumulant generating function becomes

$$K(t) \;=\; \bar{\xi}^{-1} \sum_{n=1}^{\infty} \frac{S_n}{n!}\,(\bar{\xi}t)^n\;. \tag{38}$$

The relevance of the hierarchical scaling lies in the fact that it is supported by observations (see below) and that it finds dynamical justifications in the framework of the gravitational instability picture.

In the mildly non–linear regime of gravitational clustering, hierarchical scaling is predicted by perturbative approaches. Peebles [58] showed that, applying second–order perturbation theory to the unsmoothed density field, it turns out that $S_3 =$ 34/7. Fry [24] demonstrated that hierarchical correlations of any order follow from perturbative analysis. Juszkievicz et al. [37] found that $S_3$ depends on the spectrum profile and, for a top–hat window, it is

$$S_3 = \frac{34}{7} - (n+3)\,, \tag{39}$$

where $n$ is the spectral index, $P(k) \propto k^n$. Bernardeau [4] developed a general formalism to work out the expression of $S_N$ at the generic order $N$, still for a top–hat window. Catelan & Moscardini [13] and Lokas et al. [47] provided expressions for $S_3$ and $S_4$ in the case of Gaussian window.

In the strongly non–linear regime, the hierarchical scaling is predicted by the closure of the BBGKY equations [23, 24, 35], although no general agreement exists between different authors about the sequence of the $Q_n$ coefficients.

## 3. Error analysis

One of the most important issues in any statistical analysis of the galaxy and cluster distributions is related to the estimate of the uncertainties that should be attached to the measured quantities. Several sources of errors are in general present, which are connected mainly to the limited number of objects included in any observational sample and to the finite size of the sampled portion of the Universe.

Taking properly into account such uncertainties is of crucial relevance for at least two reasons: **(a)** to establish the statistical significance of any measured clustering signal; **(b)** to assess by how much this signal is different with respect to that provided by a reference model, such as a cosmological simulation of a given dark matter scenario.

### *3.1. Sampling errors*

They are due to the finite number of points in a given data set, whose effect is that of producing a noisy sampling of the underlying statistics that one would measure. A first prescription to estimate such errors relies on the assumption that the observed object distribution is a Poissonian sampling of an underlying statistics. In this way, the relative statistical uncertainty of a given measure follows the $1/\sqrt{N}$ rule, $N$ being the number of "data" on which that measure is based. As an example, let us consider the two–point correlation function, $\xi(r)$, which can be estimated as

$$\xi(r) = \frac{DD(r)}{RR(r)} - 1 \tag{40}$$

(cf. ref.[49], and references therein). In the above relation $DD(r)$ is the number of pairs of data points at separation $r$, while $RR(r)$ is the number of pairs for a random distribution having the same number of points as in the real one. Accordingly, the expected uncertainty in the $RR$ determination is $\sigma_{RR} = \sqrt{RR}$ and corresponds to the scatter between different realizations of the random sample. Therefore, the Poissonian error in the estimate of $\xi(r)$ is

$$\sigma_\xi(r) = \sqrt{\frac{1+\xi(r)}{RR(r)}} = \frac{1+\xi(r)}{\sqrt{DD(r)}}\,, \tag{41}$$

which has the expected scaling with $DD$. However, as we also discussed for the shot–noise corrections of eqs.(33), the assumption of Poissonian sampling of a background statistics is always rather problematic.

In order to have more realistic estimates of the sampling errors, a possibility is to find a suitable way to slightly perturb the point distribution and check the stability of the statistical measure against the perturbation. A vastly applied method to perturb the original sample is based on the so–called bootstrap resampling technique (see, e.g., ref.[46]). This method is based on the generation of pseudo data sets, that are obtained by randomly selecting $N$ data points from the original data set containing as many points, by allowing for repetition. More in detail, suppose that $\mathbf{X} = \{X_1, \ldots, X_N\}$ is the set of $N$ raw data. A bootstrap sample $\mathbf{Y}$ is then obtained by randomly sampling the $\mathbf{X}$ vector $N$ times. By repeating this operation $n$ times, one ends up with an ensemble of $\mathbf{Y_i}$ $(i = 1, \ldots, n)$ of bootstrap samples. If $w_i^*$ is the result of some measure on the $i$–th bootstrap sample (for example, the two–point correlation function $\xi(r)$), then the variance over the $\mathbf{Y}_i$ ensemble is

$$\sigma_n^2 = \frac{\sum_{i=1}^{n}(w_i^* - \bar{w}^*)^2}{n-1}, \qquad (42)$$

with $\bar{w}^* = \sum_{i=1}^{n} w_i^*/n$ the bootstrap–averaged quantity. Under general conditions, it can be proved [40] that the variance evaluated over the ensemble of such bootstrap resamplings converges to the true sampling error when the number of such resamplings is sufficiently large, $\sigma_{true}^2 = \lim_{n\to\infty} \sigma_n^2$. It is also possible to show [54] that in typical cases the bootstrap errors are about a factor $\sqrt{3}$ larger than the Poisson error of eq.(41).

It is worth recalling that the bootstrap resampling procedure gives only the sampling uncertainties in the estimate of the $w$ quantity, whose value must not be confused with the bootstrap–averaged $\bar{w}^*$, which can be in general different since it refers to the perturbed data set.

### *3.2. Cosmic variance*

This kind of uncertainty arises because of the intrinsically limited extension of the volume surveyed by any sample. In fact, we expect that different portions of the Universe differ from each other, the amount of the scatter being rather large for small patches, while decreasing for larger and larger volumes, as the "fair sample" size is approached.

On the other hand, typical sizes of available observational samples are by far not big enough to allow an estimate of the large–scale cosmic variance over a sufficiently large number of independent volumes. In order to overcome this limitation one resorts to numerical simulations to estimate the amount of cosmic variance. It is however clear that its value depends on the assumed cosmological model. In particular, a larger scatter is expected if the assumed model power spectrum has larger fluctuations on large scale.

Therefore, the best strategy to compare data and theoretical models by including the effects of cosmic variance can be sketched as follows.

(a) Run for each model a large number of independent realizations, each reproducing the basic features (i.e., object number density and selection functions) of the real data set.

(b) Check how many of such realizations reproduce the results of the real data analysis, so as to assess the reliability of that model.

The effect of the cosmic variance on the clustering analysis of galaxy clusters can be judged from Figure 2, where the heavy crosses corresponds to the bootstrap uncertainties in the variance and skewness analysis (see below) while the clouds of points are due to the cosmic variance effect for each of the six simulated dark matter models. In this plot each open symbol corresponds to one realization of the Abell/ACO observational sample for different dark matter models. It is interesting to note that different realizations can provide results which may differ from each other by a rather large amount; at the smoothing scale $R_{sm} = 30\,h^{-1}$Mpc, the scatter is of about half an order of magnitude for the variance $\sigma^2$ and about one order of magnitude for the skewness $\gamma$. This suggests that it can be quite dangerous to draw conclusions about the reliability of a dark matter model only on the ground of few realizations.

## 4. Results from observations

### *4.1. Correlation analysis*

Analyses of the three– and four–point correlation functions for galaxy and cluster distributions have been pursued by several authors from about fifteen years using both angular and redshift samples (see refs.[8, 69] for the relevant references).

As for galaxies, results converge to indicate that the hierarchical model of eq.(35) is a good fit to data with $Q \simeq 1$. More recently, several attempts have been pursued to work out the reduced cumulant $S_n$ from different observational samples. The general outcome is that the hierarchical scaling is rather well reproduced. As an example, I plot in Figure 1 the results of the $S_3$ and $S_4$ analysis for a volume–limited subsample of the Perseus–Pisces Survey (PPS) [31], which contains all the galaxies with absolute magnitude $M < -19$. The analysis has been realized by computing the moments of counts

$$\langle N^n \rangle = \frac{1}{M} \sum_{i=1}^{M} N_i^n \,, \tag{43}$$

where $N_i$ is the number of objects within the $i$–th sphere ($i = 1, \ldots, M$), which is randomly placed within the sample boundaries. In the present analysis, at each scale $R$ the total number of sampling spheres is $M = 2V_{tot}/V(R)$, where $V_{tot}$ is the total sample volume, $V(R) = (4\pi/3)R^3$ and the factor two should account for the presence of clustering in the galaxy distribution. The plotted errorbars are the r.m.s. bootstrap scatter estimated over 40 resamplings, which have been checked to be enough to ensure the convergence of the bootstrap procedure.

The dashed lines correspond to the best–fitting values of the reduced cumulants, $S_3 = 1.8$ and $S_4 = 4.8$. If one would compare the above $S_3$ value with the prediction (39) of perturbation theory, it turns out that $n \simeq 0$ for the effective spectral index at the scales considered here ($\lesssim 10\,h^{-1}$Mpc). This value seems quite large if compared to the predictions of current dark matter models, like CDM or CHDM, at the same scales. However, there are several reasons for caution when doing this kind of comparisons:

**(a)** eq.(39) holds only in the dynamical regime where perturbation theory is expected to hold, that is $k_2 = \bar{\xi} \lesssim 1$;

**(b)** effects of redshift–space distortions can significantly affect the clustering pattern in observational samples;

**(c)** eq.(39) refer to the whole matter density field, while galaxies are expected to be biased and point–like tracers of this continuous distribution.

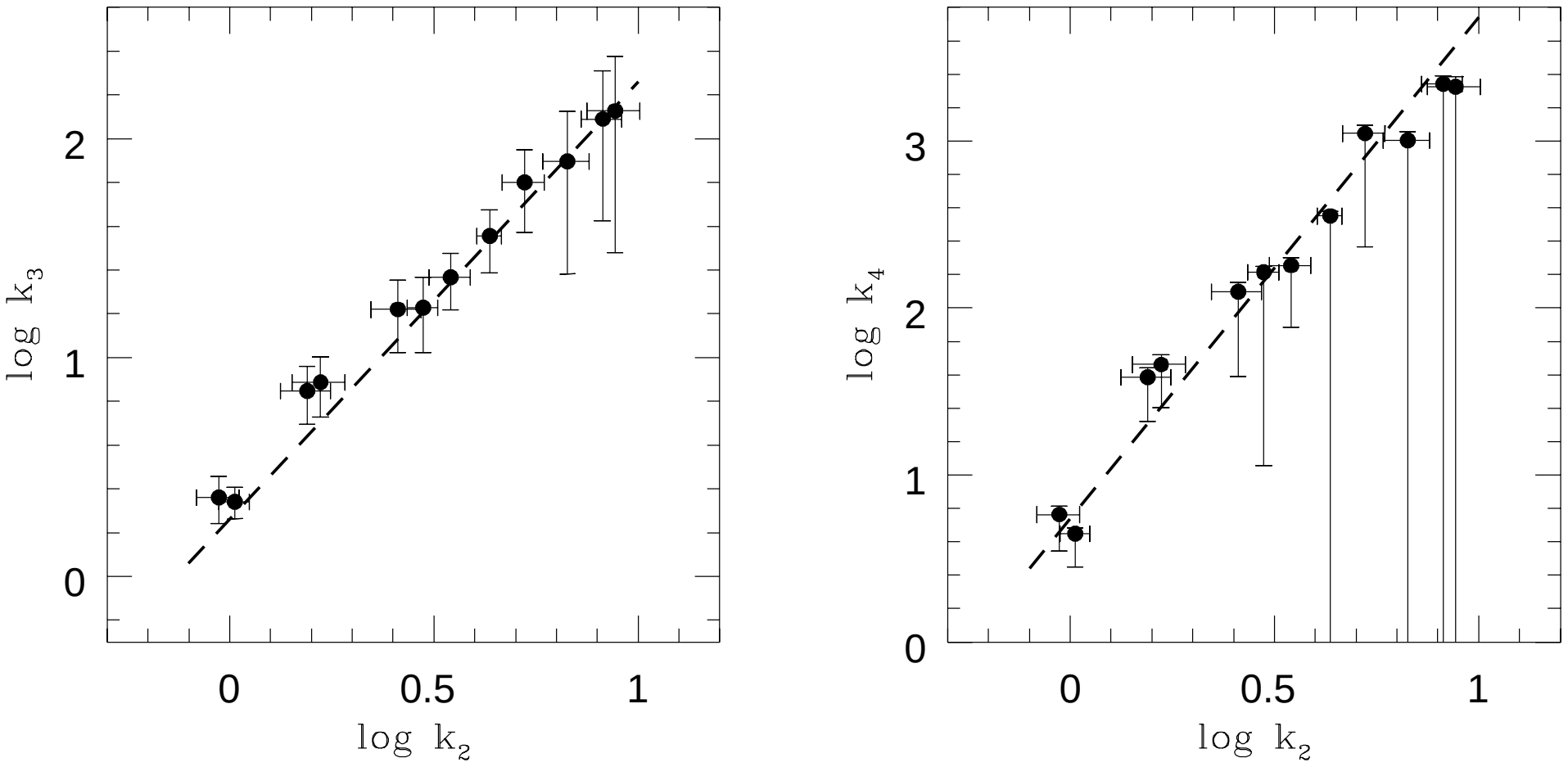

**Figure 1.** The variance–skewness (left panel) and the variance–kurtosis (right panel) relations for a volume–limited subsample of the Perseus–Pisces redshift survey. The dashed lines represent the best fitting hierarchical predictions (see Table 1).

As a matter of fact, the non–linear clustering developed by N–body simulations displays in general a much less accurate hierarchical scaling than observations (see, e.g., refs.[71, 48]). This behaviour has been interpreted in terms of sampling limitations [20], redshift–space distortions, which could make $S_n$ more constant in redshift– than in real–space [45, 70, 33] (see however ref.[27]), and high–peak identification for galaxies [7].

For these reasons, the best way to compare model predictions to observations is to pick up galaxies from N–body simulations in some realistic (physical) way and extract mock samples by reproducing as best as we can the observational biases (i.e., sample boundaries, selection functions, object number density, etc; see also ref.[64]). In Table 1 I compare the results for the PPS sample to those for mock samples extracted from high–resolution N–body simulations for a Cold+Hot Dark Matter (CHDM) model, which contains 30% of hot component contributed by one massive neutrino (see, e.g., ref.[42] for the cosmological relevance of this model). Also reported are the corresponding values taken from the literature for other samples. Apart from the remarkable agreement between real and simulated PPSs, we note that all the results converge to indicate that $S_3 \simeq 2$ and $S_4$ in the range 5–10 characterize the galaxy clustering.

As for clusters, despite the lower significance of the signal with respect to galaxies, due to the sparseness of the distribution, there is evidence that their three–point correlation function agrees with the hierarchical model for $Q \simeq 0.6$ (see, e.g., ref.[8] and references therein). As for the count–in–cell statistics, several authors found recently that $S_3 \simeq 2$ for both angular [9] and redshift [60, 29] samples.

In Figure 2 I report the variance–skewness relation for Abell/ACO clusters and compare it with similar results from the analysis of an extended set of cluster simulations based on an optimized implementation of the Zel'dovich approximation [68], as described in ref.[59]. The moments for the cluster distribution have been estimated by using a Gaussian window and the two plotted data refer to $R = 20$ and $30\,h^{-1}$Mpc for the window radius. The six panels are for different dark matter models and each point refers to one simulation of that model, containing about the same number of points as the real sample. As already mentioned, the scatter of such points gives an idea of the cosmic variance, i.e. of the variation of the results when taking different patches of the Universe. Apart from the details of the dark matter models (see ref.[10]), it is worth noting how discriminatory this statistic is; for several models almost no observer measures cumulants in the observational range, while for other models the real data are rather typical.

**Table 1.** Values of the $S_3$ and $S_4$ coefficients for simulations and for real galaxy samples. Only results for PPS, for IRAS by Bouchet et al. [11] and for CHDM simulations refer to redshift space.

| Sample | $S_3$ | $S_4$ |
|---|---|---|
| CHDM | $1.8 \pm 0.3$ | $6.1 \pm 2.2$ |
| PPS | $1.8 \pm 0.2$ | $4.8 \pm 1.5$ |
| CfA [58] | $2.4 \pm 0.2$ | ...... |
| CfA [27] | $2.0 \pm 0.2$ | $6.3 \pm 1.6$ |
| SSRS [27] | $1.8 \pm 0.2$ | $5.4 \pm 0.2$ |
| IRAS [52] | $2.2 \pm 0.2$ | $10 \pm 3$ |
| IRAS [27] | $2.2 \pm 0.3$ | $9.2 \pm 3.9$ |
| IRAS [11] | $1.5 \pm 0.5$ | $4.4 \pm 3.7$ |
| APM [28] | $3.8 \pm 0.4$ | $33 \pm 7$ |

### *4.2. The probability density function*

Instead of studying the moments of a distribution, an alternative method, which is becoming increasingly popular, is the study of the probability density function (PDF). Usually one attempts to obtain a continuous density field by smoothing the discrete distribution of objects with some window function like those of eqs.(18) and (19).

Different expressions for models of the PDF have been introduced in the literature. The more common are listed as follows.

(a) The Gaussian PDF

$$p(\varrho) = \frac{1}{\sqrt{2\pi\sigma^2}} \exp\left[-\frac{(\varrho-1)^2}{2\sigma^2}\right], \tag{44}$$

where $\sigma^2$ is the variance of $\varrho = \rho/\bar{\rho}$.

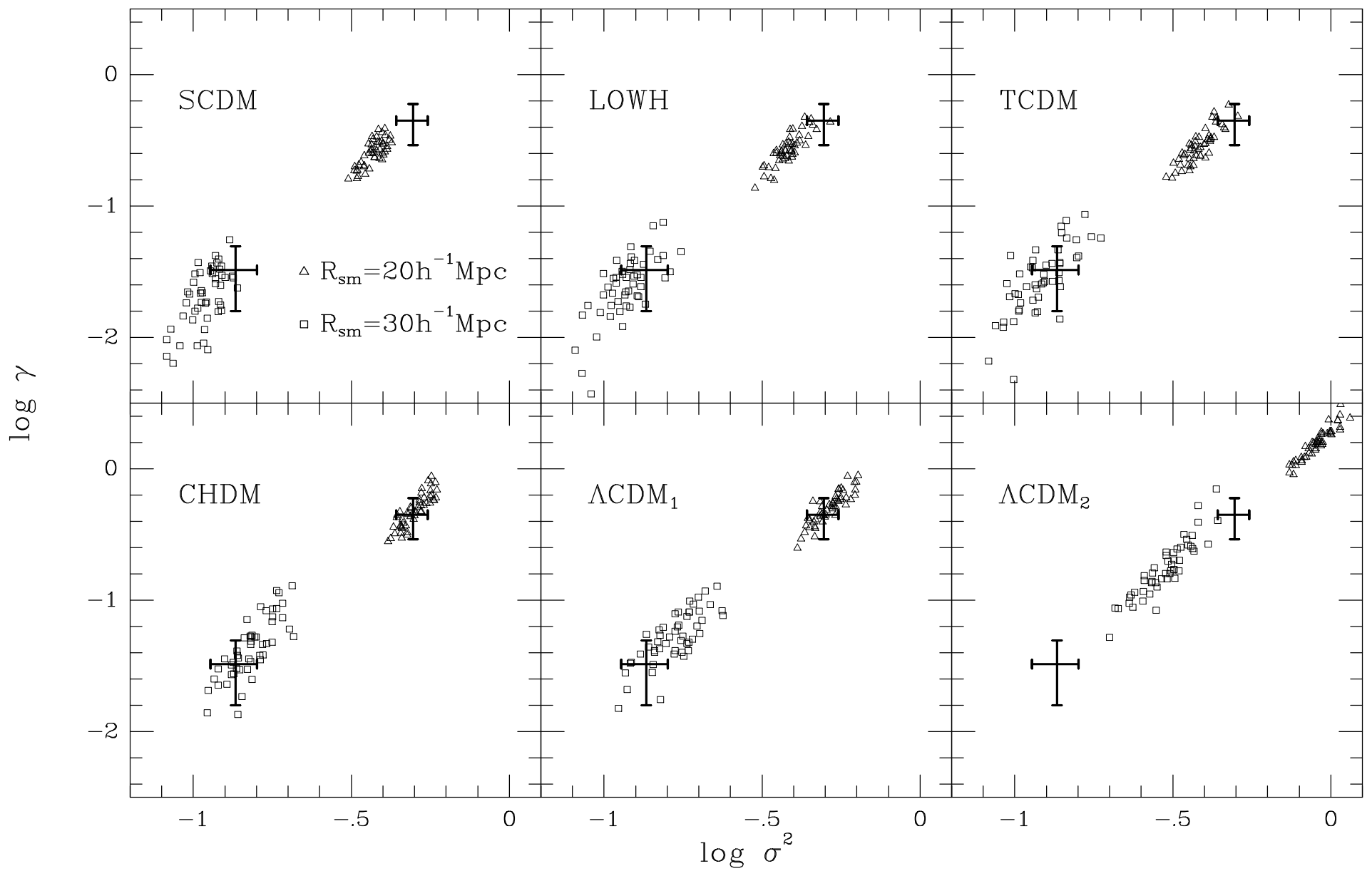

**Figure 2.** Variance–skewness relation for clusters, using the Gaussian window. The six panels are for different dark matter models Two window radii are used: $R = 20\,h^{-1}$Mpc (upper data) and $R = 30\,h^{-1}$Mpc (lower data). Heavy crosses are the observational results based on the Abell–ACO redshift cluster sample.

**(b)** The lognormal distribution given by

$$p(\varrho) = \frac{1}{\sqrt{2\pi\sigma_L^2}} \exp\left[-\frac{(\ln\varrho - \mu_L)^2}{2\sigma_L^2}\right]\frac{1}{\varrho}, \tag{45}$$

where $\varrho$ is obtained through an exponential transformation of a Gaussian random variable $\chi$ as $\varrho = \exp(\chi)$. In eq. (45), $\mu_L$ and $\sigma_L$ are the mean and standard deviation of $\chi = \ln\varrho$ respectively.

**(c)** The PDF resulting from the application of the Zel'dovich approximation to Gaussian initial fluctuations [43]:

$$\begin{aligned} p(\varrho) &= \frac{9\times 5^{3/2}}{4\pi N_s \varrho^3 \sigma^4}\int_{3\varrho^{-1/3}}^{\infty} ds\, e^{-(s-3)^2/2\sigma^2}\left(1+e^{-6s/\sigma^2}\right) \\ &\times \left(e^{-\beta_1^2/2\sigma^2} + e^{-\beta_2^2/2\sigma^2} - e^{-\beta_3^2/2\sigma^2}\right); \\ \beta_n(s) &= \sqrt{5}\,s\,\{1/2 + \cos[2/3\,(n-1)\pi \\ &+ 1/3\,\arccos\left(54/\varrho s^3 - 1\right)]\}, \end{aligned} \tag{46}$$

where $N_s$ is the average stream number per Eulerian point (see ref.[43]).

**(c)** The Edgeworth expansion [38]

$$p(\varrho) = \frac{1}{\sqrt{2\pi\sigma^2}} \exp\left[-\frac{(\varrho-1)^2}{2\sigma^2}\right] \times$$
$$\times \left[1 + \frac{S_3 H_3(x)}{6}\sigma + \left(\frac{S_4 H_4(x)}{24} + \frac{S_3^2 H_6(x)}{72}\right)\sigma^2 + \ldots\right] \quad (47)$$

where $x = \delta/\sigma$ and $H_n$ are the Hermite polynomials. This expression represents an expansion which holds for small $\sigma$ values around the Gaussian expression. Its reliability however breaks down at $\sigma \geq 0.5$, where $p(\varrho)$ becomes unphysically negative.

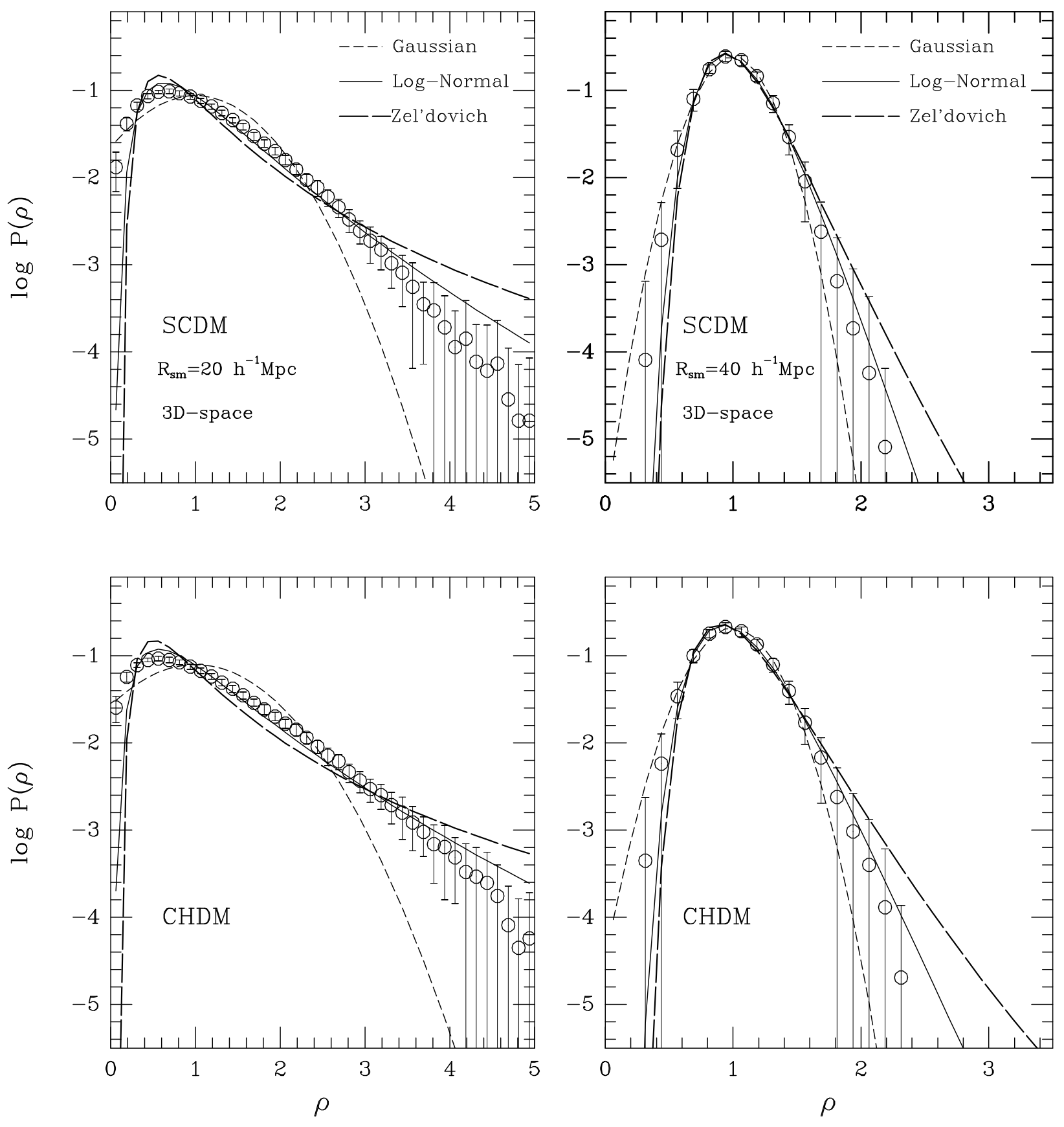

**Figure 3.** Comparison between the PDFs of simulated cluster distributions and the theoretical models, at $R_{sm} = 20$ and $40\,h^{-1}$Mpc for the Gaussian smoothing scale. Solid, long–dashed and short–dashed curves correspond to the lognormal, Zel'dovich and Gaussian model, respectively. Error bars are cosmic r.m.s. scatter evaluated over 50 realizations of each model (taken from [10].

Kofman et al. [43] computed the PDF for the IRAS sample and for the density field reconstructed with the POTENT procedure with $\Omega_0 = 1$. They found that the PDF is well modelled by the lognormal expression. By comparing these results with CDM N–body simulations, they also concluded that this results is perfectly consistent with the assumption of Gaussian initial conditions. Plionis & Valdarnini [60] and Kolatt et al. [44] studied the PDF of the Abell/ACO smoothed cluster distribution and found that also for clusters the PDF is well approximated by a lognormal distribution. In ref.[10] we compared the PDF for Abell/ACO clusters to that of cluster simulations based on six different dark matter models. We found that the shape of the PDF is a stringent tests for such models, thus suggesting its usage as a useful discriminatory statistics.

Coles & Jones [16] argued that the lognormal distribution provides a natural description for density perturbations resulting from Gaussian initial conditions in the weakly non–linear regime. On the other hand, Bernardeau & Kofman [5] have shown that the lognormal distribution is not really a natural consequence of mildly non–linear gravitational evolution, but a very convenient fit only in some portion of the $(\bar{\xi}, n)$–plane (i.e. $\bar{\xi} \ll 1$ and spectral index $n \approx -1$).

In Figure 3 I plot the PDF for cluster simulations. The plots refer to simulations based on two rather different initial spectra, namely the standard CDM model (top panels) and the CHDM model (bottom panels). It turns out that, despite the difference between the two initial spectra and the fact that the underlying dynamics is regulated by the Zel'dovich approximation, the lognormal expression fares much better than that of eq.(46). Note also that the lognormal model remains a better fit than the Gaussian one also at the larger smoothing scale, where the variance is well below unity ($\sigma^2 \simeq 0.06$ and $\sigma^2 \simeq 0.04$ for SCDM and CHDM, respectively; see ref.[10]).

## 5. Geometrical descriptions of the LSS

The variety of structures in the galaxy distribution, like filaments, voids, clusters, extending over a broad range of scales, calls for a global description of the geometry of the LSS. Although the correlation analysis provides rather useful information, nevertheless it says only a little about the "shape" of the galaxy distribution. For this reason, many attempts have been devoted to develop and apply statistical methods, which were able to provide such a description. A treatment of all such approaches is beyond the scope of these *Lectures.* They include the study of percolation properties [41], Minkowski functionals (see ref.[67] and references therein), structure of the minimal spanning tree [6] or other graph statistics, filamentarity analyses (see, e.g., ref.[21]), etc. In the following I will only describe two popular examples of such statistics, namely the void probability function (VPF) and the topology analysis of the genus characteristics.

### *5.1. The void probability function*

The void probability function (VPF) is defined as the probability of finding no objects within randomly placed sampling volumes. According to its definition, it represents the $N = 0$ case of the $P_N$ count probabilities defined by eq.(31). Therefore, it is connected to the sequence of cumulants as

$$P_0(R) = M(-\bar{N}) \;=\; \exp\left[\sum_{n=1}^{\infty} \frac{(-\bar{N})^n}{n!}\, k_n(R)\right] \tag{48}$$

($k_1(R) \equiv 1$). Since $P_0$ conveys information about correlations of any order, the VPF statistic has been suggested as a useful tool to provide a global clustering characterization. Note, however, that $P_0$ depends only on the number of non–empty cells, with no regard to the number of objects contained inside them. For this reason, it provides only a description of the geometry, rather than of the clustering, of a point distribution.

For a completely uncorrelated (i.e. Poissonian) distribution, it is $P_0 = \exp(-\bar{N})$, so that any departure of the quantity

$$\sigma(\bar{N}, R) \;=\; \frac{-\log(P_0)}{\bar{N}} \tag{49}$$

from unity represents the signature for the presence of clustering.

Assuming hierarchical scaling for correlations, and owing to the expression (37) for $k_n$, it follows that

$$\sigma(N_c) \;=\; \sum_{n=1}^{\infty} \frac{(-N_c)^{n-1}}{n!} S_n \,, \tag{50}$$

where $N_c = \bar{N}\bar{\xi}$ is the average object count in excess with respect to a random distribution. Therefore, while the value taken by $\sigma - 1$ states the deviation of the distribution from Poisson, the scale dependence of $\sigma$ in the hierarchical scaling regime can be expressed directly through $N_c$ (and not through $\bar{N}$ and $R$ separately).

In the analysis of their scale-invariant model for correlation functions,

$$\kappa_n(\lambda \mathbf{x}_1, \ldots, \lambda \mathbf{x}_n) \;=\; \lambda^{-\gamma(n-1)} \kappa_n(\mathbf{x}_1, \ldots, \mathbf{x}_n) \tag{51}$$

(where $\bar{\xi}(R) \propto R^{-\gamma}$), Balian & Schaeffer [2] found that for asymptotically large $N_c$ the power-law relation $\sigma(N_c) \propto N_c^{-\omega}$ should hold, with $0 < \omega < 1$. In the framework of hierarchical correlation pattern, several models have been proposed, each providing a different expression for the VPF. Among these models is the thermodynamical one [66], which predicts

$$\sigma(N_c) \;=\; (1 + N_c)^{-1/2} \,. \tag{52}$$

A further model [26] describes the galaxy clustering as due to a Poissonian distribution of clusters, each containing a suitable number of members. The resulting hierarchical Poisson distribution gives

$$\sigma(N_c) \;=\; \frac{1 - e^{-N_c}}{N_c} \,. \tag{53}$$

The negative binomial model [12] predicts

$$\sigma(N_c) \;=\; \frac{\log(1 + N_c)}{N_c} \tag{54}$$

and has been shown to provide a quite good fit to CfA data [30]. Finally, the phenomenological model

$$\sigma(N_c) \;=\; \left(1 + \frac{N_c}{2\omega}\right)^{-\omega} \tag{55}$$

has been proposed by Alimi et al. [1], which found a best fit to the CfA data for $\omega = 0.50 \pm 0.15$ (note that for $\omega = 0.5$ eq.[55] coincides with the thermodynamical model). A similar result has also been found by Maurogordato et al. [51] from the analysis of the SSRS survey and by Bonometto et al. [7] from the analysis of CDM and CHDM simulations.

The VPF has been suggested as a potentially powerful discriminant between different cosmological models. It is however clear that it is also very sensitive to the details of the object distribution. For instance, adding few points in underdense regions is expected not to significantly affect correlation functions, while it may greatly modify the VPF. Indeed, Weinberg & Cole [73] found that the VPF is sensitive to the galaxy identification scheme in N–body simulations. In addition, selection effects in real samples, like boundary geometry, redshift–space distortions, etc., can make rather difficult any comparison of VPF results for real and simulated universes.

In Figure 4 I report the VPF results of the comparison between CDM and CHDM simulations and a volume–limited subsample of the Perseus–Pisces redshift survey [32]. The four panels are for two different realizations of CHDM and for two different evolutionary stages of CDM. The dashed curves represent the shot–noise level, $P_0 = e^{-\bar{N}}$, the solid curve is for the real data set and the different dotted curves in each panel are for different mock samples extracted from each simulation box. Note that in this case different samples do not involve independent volumes. Instead, they are obtained by sampling almost the same simulation volume from different advantage points. Therefore, the scatter between the curves is the effect of a kind of "local variance", rather than of the cosmic variance, whose larger effect may be judged by comparing the results for the two CHDM realizations.

It is remarkable that differences between different realizations are of the same order as differences between different dark matter models. Therefore, although such results confirm the VPF as a useful discriminatory statistics, they also suggest the necessity of having cosmic–variance effects under control, either choosing a larger box, or running constrained simulations, which contain *ab initio* the essential features of a specific observational sample.

### *5.2. Topology*

Instead of providing a detailed description of topological concepts in a formal mathematical language, in the following I only briefly introduce the measures of topology which are applied in cosmological context and what we learn from their application (see ref.[53] for a review about the cosmological applications of topology measures). In this context, the concept of "genus" has been introduced to describe the topology of isodensity surfaces, drawn from a density field. The genus $G$ of a surface can be defined as

$$G \;=\; (\text{number of holes}) - (\text{number of isolated regions}) + 1\,. \qquad (56)$$

Therefore, a single sphere has genus $G = 0$, a distribution made of $N$ disjoint spheres has $G = -(N-1)$, while $G = 1$ for a torus. More in general, the genus of a surface corresponds to the number of "handles" it has, or, equivalently, to the number of cuts that can be realized on that surface without disconnecting it into separate parts. A more formal definition of genus can be given by means of the Gauss–Bonnet theorem (see, e.g., ref.[56]), which relates the curvature of the surface to the number of holes. According to this theorem, for any compact two–dimensional surface the genus $G$ is

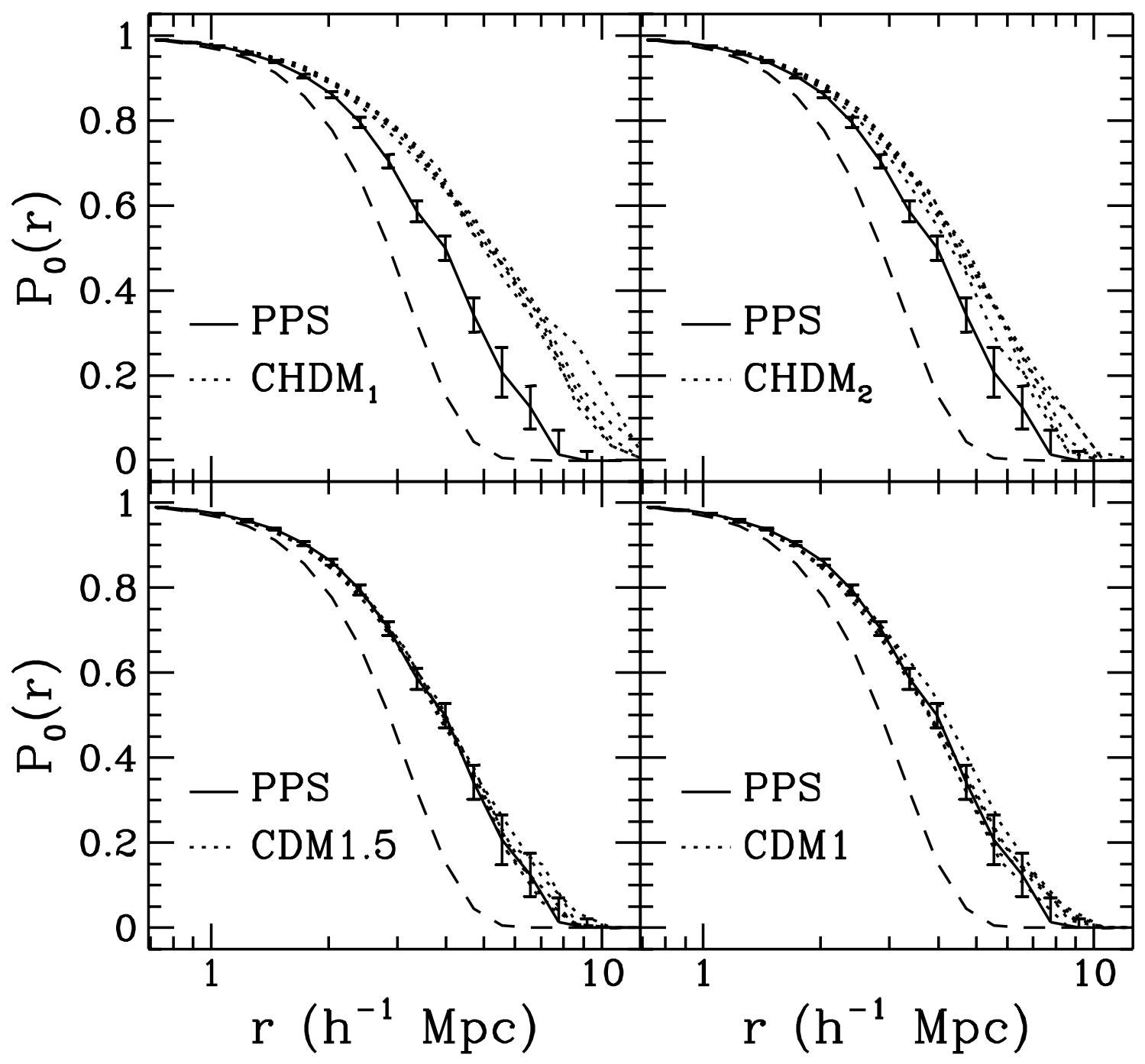

**Figure 4.** The scale–dependance of the VPF $P_0(R)$ is shown for the $M_{\rm lim} < 19$ volume–limited sample of the Perseus–Pisces Survey (continuous curve) and for five different artificial VLS's obtained from each simulation (dotted curves). The five realizations of artificial VLS's have different observer positions but the same number of galaxies as in the real VLS. The dashed curve is what one expects for a Poissonian distribution (taken from [33]).

related to the curvature $C$ according to

$$C \;=\; \int K\,dA \;=\; 4\pi(1-G)\,. \tag{57}$$

Here $K$ represents the local Gaussian curvature of the surface that, at each point, is defined as the reciprocal of the product of the two principal curvature radii, $K=(a_1a_2)^{-1}$. Since $K$ has the dimension of length$^{-2}$, the curvature $C$ and, thus, the genus are dimensionless quantities. For a sphere of radius $r$ it is $K=r^{-2}$, so that $C=4\pi$ and $G=0$, as previously argued. Strictly speaking, while the genus of a surface gives the number of its "handles", eq.(56) defines a related quantity, that is the Euler–Poincaré (EP) characteristic [56]. In a sense, we can say that, while the genus deals with the properties of a surface, the EP characteristics describe the properties of the excursion set, i.e. of the part of the density field exceeding a density threshold value. Based on the Gauss–Bonnet theorem, it can be proved that genus and EP characteristics are completely equivalent in the three–dimensional case.

In order to quantify the genus of the observed large scale clustering, the first step is to extract a continuous density field starting from the discrete object distribution. This can be done by collecting the points in cells and then by smoothing the resulting cell count with a suitable window function. In order to keep Poissonian shot–noise from dominating the geometry of the smoothed field, the smoothing radius should be chosen not to be much smaller than the typical correlation length.

In topology analysis it is useful to study the dependence of the genus of isodensity surfaces on the value of the density thresholds. If a high density value is selected, only few very dense and isolated regions will be above the threshold and the genus is negative. For a very low threshold, only few isolated voids are identified and, again, the corresponding genus is negative. For thresholds around the median density value we expect in general that the isodensity surfaces have a multiply connected structure, with a resulting positive genus. These general considerations can be verified on a more quantitative ground for models having an analytical eexpression for the genus. The simplest case occurs for a Gaussian random field (see, e.g., ref.[3]), which, in three dimensions, has a threshold–dependent genus per unit volume

$$g(\nu) \;=\; \frac{1}{(2\pi)^2}\left(\frac{\langle k^2\rangle}{3}\right)^{3/2}(1-\nu^2)\,e^{-\nu^2/2}\,. \qquad (58)$$

The density threshold is set so as to select only fluctuations exceeding $\nu$ times the r.m.s. value $\sigma$. Therefore, $g(\nu)$ describes the topology of the isodensity surfaces, where the fluctuations take the value $\delta = \nu\sigma$. Moreover,

$$\langle k^2\rangle \;=\; \frac{\int P(k)\,W^2(k)\,k^2\,d^3k}{\int P(k)\,W^2(k)\,d^3k} \qquad (59)$$

is the second order spectral moment, which depends on the choice of the window $W(k)$ used to smooth the discrete distribution. According to eq.(59), $g(\nu)$ depends on the shape of the power spectrum, but not on its normalization. Since the amplitude of the genus curve turns out to depend on the profile of the power spectrum through the second–order spectral moment, repeating genus measures for different smoothing radii gives information about the shape of $P(k)$. Following eq.(58), several interesting features of the $g(\nu)$ curve appear. First of all, as expected for a Gaussian field, which has the same structure in the overdense and underdense regions, $g(\nu)$ is an even function of $\nu$, with its maximum at $\nu = 0$. This is characteristic of the so–called "sponge–like" topology. For $|\nu| < 1$ it is $g(\nu) > 0$, due to the multiple connectivity of the isodensity surfaces, while $g(\nu) < 0$ for $|\nu| > 1$, due to the predominance of isolated clusters. Different topologies are however expected when non–Gaussian fields are considered [14].

In the case of a distribution realized by superimposing dense clusters on a smooth background, isolated structures start dominating also at rather low density values and the $g(\nu)$ curve peaks at negative $\nu$'s. Vice versa, at large and positive $\nu$ values the distribution is that of isolated regions and $g(\nu)$ becomes more negative than expected for a Gaussian field. This case is usually referred to as "meatball" topology. The opposite case occurs when the distribution is dominated by big voids, with objects arranged in sheets surrounding the voids. The resulting topology is usually called "cellular" or "Swiss–cheese" and the corresponding $g(\nu)$ curve peaks at positive $\nu$'s.

Topological measures can also be usefully employed when dealing with two–dimensional density fields. However, in this case some ambiguities arise, for example in distinguishing whether an underdense area is due to a tunnel or to a spherical void in three–dimensions. In addition, the interpretation of the genus in terms of the number of handles of an isodensity surface can not be applied in two dimensions. In this case, the topology measure is represented by the EP characteristics, which is defined as the difference between the number of isolated high–density regions and the number of isolated low–density regions. The EP characteristics per unit area at the overdensity level $\nu$ for a Gaussian random field is

$$g(\nu) \;=\; \frac{1}{(2\pi)^{3/2}}\,\frac{\langle k^2\rangle}{2}\,\nu\,e^{-\nu^2/2}\,, \tag{60}$$

so that $g(\nu)$ is an odd function of $\nu$ and $g(0)=0$.

Application of the genus statistics to the study of LSS has been employed in recent years, analysing both the evolution of N–body simulations and observational data sets. Measures of the EP characteristics for the angular galaxy distribution [19] and of the genus for three–dimensional redshift surveys [34, 55, 72] consistently shows a too large genus amplitude if compared to N–body simulations of the standard CDM model. Furthermore, all the analyses indicate the presence of a slight meatball shift at small smoothing scales, followed by a sponge–like topology at larger scales, as expected on the ground of Gaussian initial fluctuations. Although the meatball shift at small scales is expected on the ground of non–linear gravitational evolution [50], attempts have been also devoted to check whether this result implies non–Gaussian initial conditions for the CDM model [18].

The application of the same analysis to galaxy clusters has been also realized both for their projected distribution [61] and for redshift surveys [65]. Also in this case, a slight meatball shift is observed, which is however consistent with expectations based on random–phase initial conditions. Although the genus for the galaxy distribution has been compared quite in detail with simulations based on different dark matter models [72], the same has not yet been realized for clusters. This point will surely deserve future investigations.

## 6. Conclusions

As already anticipated in the Introduction, these *Lectures* should not be considered a comprehensive review about the application of advanced statistical methods in cosmology, for at least two reasons. Firstly, I gave only a partial view of the many techniques, which have been applied until now to characterize the galaxy clustering. My aim has been to introduce different statistical concepts which are able to pick up different characteristics of the large–scale structure (e.g., correlation properties, geometry and topology). However, one can well imagine other measures, like fractal scaling, filamentarity and percolation, which should be considered as complementary to those I described. Secondly, I dealt here with the characterization of the large–scale structure only in configuration space, while no words have been spent about the statistics of the velocity field traced by cosmic structures. This represents a relatively more recent field, which has undergone a progressive development during the last few years, thanks to the availability of more and more reliable redshift–independent measurements of galaxy distances (see refs.[22, 69] for recent reviews).

However, even relying on the material I presented here, at least two firm points can be established.

**(a)** Today available samples of galaxies and galaxy clusters are already large enough to provide reliable clustering information, which go beyond the 2–point correlation statistics.

**(b)** Such measures are discriminatory, in the sense that they often allow to distinguish between different dark matter models at a quite high confidence level.

It is however clear that we are probably still far from having reached a satisfactory and self–consistent understanding of the formation and evolution of cosmic structures on the ground of the analysis of their distribution. However, we are at a point in which we expect in the reasonably near future a better clarification of both theoretical and observational aspects concerning the large–scale structure. Hopefully this will allow us to further restrict the range of viable models or will lead to a radical change in our view of the Universe. From the theoretical side, a crucial point concerns a deeper understanding of the physics underlying galaxy formation. One's hope is to address adequately this problem with the availability of new numerical techniques and computing facilities, so as to make clear what we are comparing to what when analyzing numerical simulations and observational data. From the observational side, we are waiting for the advent of new huge galaxy redshift surveys (e.g., SDSS), as well as the compilation of new cluster samples (e.g., ROSAT).

For these reasons, I believe that the study of the large–scale structure will remain in the following years an exciting field of investigation: the development and refinement of methods of statistical analysis will represent a necessary ingredient to clarify our view of the Universe.

**Acknowledgments.**

I wish to acknowledge Joel Primack for useful suggestions about the presentations of this material. Many thanks are also due to Paolo Tini Brunozzi for a careful reading of a preliminary version of the manuscript.

## Appendix A. The functional derivative

In order to introduce the concept of functional derivative of a given functional $F[\tau(\mathbf{x})]$ with respect to $\tau(\mathbf{x}) \in \mathcal{F}$, let us consider a small function $\delta\tau(\mathbf{x}) \in \mathcal{F}$, so that $\tau(\mathbf{x}) + \delta\tau(\mathbf{x})$ differs from $\tau(\mathbf{x})$ only in a neighbourhood of $\mathbf{x} = \mathbf{y}$. Moreover, let

$$\delta\omega = \int d\mathbf{x}\, \delta\tau(\mathbf{x}) \tag{A1}$$

be the volume element in $\mathcal{F}$ contained between $\tau(\mathbf{x})$ and $\tau(\mathbf{x}) + \delta\tau(\mathbf{x})$.

The functional derivative of $F[\tau]$ is defined as

$$\frac{\delta F}{\delta\tau(\mathbf{y})} = \lim_{\delta\omega\to 0} \frac{F[\tau+\delta\tau] - F[\tau]}{\delta\omega}\,. \tag{A2}$$

Taking $\delta\tau(\mathbf{y}) = \delta\omega\,\delta(\mathbf{x}-\mathbf{y})$, eq.(A2) becomes

$$\frac{\delta F}{\delta\tau(\mathbf{y})} = \lim_{\delta\omega\to 0}\frac{F[\tau(\mathbf{x})+\delta\omega\,\delta(\mathbf{x}-\mathbf{y})]-F[\tau(\mathbf{x})]}{\delta\omega} \qquad \text{(A3)}$$

that, in the particular case $F[\tau(\mathbf{x})] = \tau(\mathbf{x})$, reads

$$\frac{\delta\tau(\mathbf{x})}{\delta\tau(\mathbf{y})} = \delta(\mathbf{x}-\mathbf{y})\,. \qquad \text{(A4)}$$

In a similar way, higher order functional derivatives can be introduced.

As an example, let us consider the functional

$$F_q[\tau] = \int d\mathbf{x}_1 \ldots \int d\mathbf{x}_q\, f(\mathbf{x}_1,\ldots,\mathbf{x}_q)\,\tau(\mathbf{x}_1)\ldots\tau(\mathbf{x}_q)\,, \qquad \text{(A5)}$$

where $f(\mathbf{x}_1,\ldots,\mathbf{x}_q)$ is a symmetric function with respect to the variables $\mathbf{x}_1,\ldots,\mathbf{x}_q$. Differentiating the functional (A5), we get

$$\frac{\delta F_q}{\delta\tau(\mathbf{y})} = \int d\mathbf{x}_2 \ldots \int d\mathbf{x}_q\, f(\mathbf{y},\mathbf{x}_2,\ldots,\mathbf{x}_q)\,\tau(\mathbf{x}_2)\ldots\tau(\mathbf{x}_q) + \ldots$$

$$\ldots + \int d\mathbf{x}_1 \ldots \int d\mathbf{x}_{q-1}\, f(\mathbf{x}_1,\ldots,\mathbf{x}_{q-1},\mathbf{y})\,\tau(\mathbf{x}_1)\ldots\tau(\mathbf{x}_{q-1})\,. \qquad \text{(A6)}$$

Making use of the symmetry of $f(\mathbf{x}_1,\ldots,\mathbf{x}_q)$ and relabeling the integration variables, we finally obtain

$$\frac{\delta F_q}{\delta\tau(\mathbf{y})} = q\int d\mathbf{x}_1 \ldots \int d\mathbf{x}_{q-1}\, f(\mathbf{x}_1,\ldots,\mathbf{x}_{q-1},\mathbf{y})\,\tau(\mathbf{x}_1)\ldots\tau(\mathbf{x}_{q-1})\,. \text{(A7)}$$

Similarly, the higher order derivatives are

$$\frac{\delta^n F_q}{\delta\tau(\mathbf{y}_1)\ldots\delta\tau(\mathbf{y}_n)} = \frac{q!}{(q-n)!}\int d\mathbf{x}_1 \ldots \int d\mathbf{x}_{q-n}\,\times$$
$$f(\mathbf{x}_1,\ldots,\mathbf{x}_{q-n},\mathbf{y}_1,\ldots,\mathbf{y}_n)\,\tau(\mathbf{x}_1)\ldots\tau(\mathbf{x}_{q-n}) \qquad (n\le q)$$
$$\frac{\delta^n F_q}{\delta\tau(\mathbf{y}_1)\ldots\delta\tau(\mathbf{y}_n)} = 0 \qquad (n>q)\,. \qquad \text{(A8)}$$

As a further example, let us consider the exponential functional

$$F[\tau] = \exp\left[\int d\mathbf{x}\, f(\mathbf{x})\,\tau(\mathbf{x})\right]\,, \qquad \text{(A9)}$$

whose $n$–th order derivative reads

$$\frac{\delta^n F_q}{\delta\tau(\mathbf{y}_1)\ldots\delta\tau(\mathbf{y}_n)} = f(\mathbf{y}_1)\ldots f(\mathbf{y}_n)\,\exp\left[\int d\mathbf{x}\, f(\mathbf{x})\,\tau(\mathbf{x})\right] \qquad \text{(A10)}$$

from which eq.(3) for the $n$–point correlation function follows.

## References

[1] Alimi, J.-M., Blanchard, A., Schaeffer, R., 1990, ApJ, 349, L5
[2] Balian, R., Schaeffer, R., 1989, A&A, 220, 1
[3] Bardeen, J.M., Bond, J.R., Kaiser, N., Szalay, A.S. 1986, ApJ, 304, 15

[4] Bernardeau F., 1994, A&A, 291, 697
[5] Bernardeau F., Kofman L., 1994, ApJ, 443, 479
[6] Bhavsar S.P., Ling E.N., 1988, ApJ, 331, L63
[7] Bonometto S.A., Borgani S., Ghigna S., Klypin A., Primack J.R., 1995, MNRAS, 273, 101
[8] Borgani S., 1995, Phys. Rep., 251, 1
[9] Borgani, S., Coles, P., Moscardini, L., Plionis, M. 1994, MNRAS, 266, 524
[10] Borgani S., Plionis M., Coles P., Moscardini L., 1995, MNRAS, in press, preprint ASTRO–PH/9505050
[11] Bouchet, F.R., Strauss, M.A., Davis, M., Fisher, K.B. Yahil, A., Huchra, J.P., 1993, ApJ, 417, 36
[12] Carruthers, P., Shih, C.C., 1983, Phys. Lett., 127B, 242
[13] Catelan P., Moscardini L., 1994, ApJ, 426, 14
[14] Coles P. 1988, MNRAS, 234, 509
[15] Coles P., Frenk C.S., 1991, MNRAS, 253, 727
[16] Coles P., Jones B.J.T., 1991, MNRAS, 248, 1
[17] Coles P., 1995, this volume
[18] Coles, P., Moscardini, L., Plionis, M., Lucchin, F., Matarrese, S., Messina, A., 1993, MNRAS, 260, 572
[19] Coles P., Plionis M., 1991, MNRAS, 250, 75
[20] Colombi S., Bouchet F.R., Schaeffer R., 1994, 281, 301
[21] Davé R., Hellinger D., Nolthenius R., Primack J., Klypin A., 1995, preprint ASTRO–PH/9505142
[22] Dekel A., ARA&A, 32, 99
[23] Davis, M., Peebles, P.J.E., 1977, ApJS, 35, 425
[24] Fry, J.N., 1984, ApJ, 277, L5
[25] Fry, J.N., 1984, ApJ, 279, 499
[26] Fry, J.N., 1985, ApJ, 289, 10
[27] Fry J.N., Gaztañaga E., 1993, 413, 447
[28] Gaztañaga E., 1994, MNRAS, 268, 913
[29] Gaztañaga E., Croft R.A.C., Dalton G.B., 1995, MNRAS, 276, 336
[30] Gaztañaga E., Yokoyama J., 1993, ApJ, 403, 450
[31] Giovanelli R., Haynes M.P., 1991, ARAA, 29, 499
[32] Ghigna S., Borgani S., Bonometto S.A., Guzzo L., Klypin A., Primack J.R., Giovanelli R., Haynes M.P., 1995, ApJ, 437, L71
[33] Ghigna S., Bonometto S.A., Guzzo L., Giovanelli R., Haynes M.P., Klypin A., Primack J.R., 1995, ApJ, in press
[34] Gott III, J.R., et al., 1989, ApJ, 340, 625
[35] Hamilton, A.J.S., 1988, ApJ, 332, 67
[36] Jensen, L.G., Szalay, A.S., 1986, ApJ, 305, L5
[37] Juszkievicz R., Bouchet F.R., Colombi S., 1993, 412 L9
[38] Juszkievicz R., Weinberg D., Amsterdamsky P., Chodorowski M., Bouchet F.R., 1995, 442, 39
[39] Kaiser, N., 1984, ApJ, 284, L9
[40] Kiefer J., Wolfowitz J., 1956, Ann. MAth. Stat., 27, 887
[41] Klypin A., Shandarin S.F., 1993, ApJ, 413, 48
[42] Klypin A., Holtzman J., Primack J.R., Regös E., 1993, ApJ, 416, 1
[43] Kofman L., Bertschinger E., Gelb J.M., Nusser A., Dekel A., 1994, ApJ, 420, 44
[44] Kolatt T., Dekel A., Primack J.R., 1995, in preparation
[45] Lahav O., Itoh M., Inagaki S., Suto Y., 1993, ApJ, 402, 387
[46] Ling E.N., Frenk C.S., Barrow J.D., 1986, MNRAS, 223, 21p
[47] Lokas E.L., Juszkievicz R., Weinberg D.H., Bouchet F.R., 1994, MNRAS, submitted, preprint ASTRO–PH/9407045
[48] Lucchin F., Matarrese S., Melott A., Moscardini L., 1994, ApJ, 422, 430
[49] Mart ́inez V., 1995, this volume
[50] Matsubara T., 1994, ApJ, 434, L43
[51] Maurogordato, S., Schaeffer, R., da Costa, L.N., 1992, ApJ, 390, 17
[52] Meiksin A., Szapudi I., Szalay A.S., 1992, ApJ, 394, 87
[53] Melott, A.L., 1990, Phys. Rep., 193, 1
[54] Mo H.J., Jing Y.P., Börner G., 1992, ApJ, 392, 452
[55] Moore, B., et al., 1992, MNRAS, 256, 477
[56] Nash, C., Sen, S., 1983, Topology and Geometry for Physicists (London: Academic Press)
[57] Olive K.A., 1990, Phys. Rep., 190, 307

[58] Peebles, P.J.E., 1980, The Large Scale Structure of the Universe (Princeton: Princeton University Press)
[59] Plionis M., Borgani S., Moscardini L., Coles P., 1995, ApJ, 441, L57
[60] Plionis M., Valdarnini R., 1995, MNRAS, 272, 869
[61] Plionis, M., Valdarnini, R., Coles, P., 1992, MNRAS, 258, 114
[62] Pokorski S., 1985, Gauge Field Theories (Cambridge: Cambridge University Press)
[63] Politzer, D., Wise, M., 1984, ApJ, 285, L1
[64] Primack J.R., 1995, this volume.
[65] Rhoads J.E., Gott J.R.III, Postman M., 1994, 421, 1
[66] Saslaw, W.C., Hamilton, A.J.S., 1984, ApJ, 276, 13
[67] Schmalzing, J., Kerscher M.,, Buchert T., 1995, this volume
[68] Shandarin, S.F., Zel'dovich, Ya.B., 1989, Rev. Mod. Phys., 61, 185
[69] Strauss M.A., Willick J.A., 1995, Phys. Rep., in press, preprint ASTRO–PH/9502079
[70] Suto Y., Matsubara T., 1994, ApJ, 420, 504
[71] Valdarnini, R., Borgani, S., 1991, MNRAS, 251, 575
[72] Vogeley M.S., Park C., Geller M.J., Huchra J.P., Gott J. R.III, 1994, ApJ, 420, 525
[73] Weinberg D.H., Cole S., 1992, MNRAS, 259, 652